\documentclass[fleqn,usenatbib]{mnras}

\usepackage{graphicx} 
\usepackage{color}
\usepackage{amssymb}

\def\sss{\scriptscriptstyle}
\def\U{{\sss \!U}}
\def\L{{\sss \!L}}
\def\K{{\sss \!K}}

\def\P{{\sss \!P}}
\def\S{{\sss \!S}}
\def\I{{\sss \!I}}
\def\C{{\sss \!C}}

\def\O{{\sss \!O}}

\def\nuL{\nu_\L}
\def\nuU{\nu_\U}

\def\nuISCO{\nu_{\I\S\C\O}}
\def\ISCO{\I\S\C\O}

\title[{Twin-peak QPOs and NS mass}]{{Timescale of twin-peak quasi-periodic oscillations and mass of accreting neutron stars}}
\author[T\"{o}r\"{o}k et al.]{Gabriel T\"{o}r\"{o}k$^1$\thanks{E-mail:
gabriel.torok@gmail.com}, Kate\v{r}ina Goluchov\'{a}$^1$, Eva \v{S}r\'{a}mkov\'{a}$^1$, \newauthor
Martin Urbanec$^1$, Odele Straub$^{1,2}$\\
$^1$ Institute of Physics and Research Centre for Computational Physics and Data Processing,\\
Silesian University in Opava,  Bezru\v{c}ovo n\'am.~13, CZ-746\,01 Opava, Czech Republic \\
$^2$ Max Planck Institute for extraterrestrial Physics, Gie\ss enbachstra\ss e 1, 85748 Garching, Germany
}

\begin{document}
%-------------------------------

% Dodat odkazy na hmotnosti : http://adsabs.harvard.edu/abs/2011ApJ...726...56M
%http://adsabs.harvard.edu/abs/2011ApJ...742...17L

\date{}

\pagerange{\pageref{firstpage}--\pageref{lastpage}} \pubyear{2015}

\maketitle

\label{firstpage}

%-------------------------------
\begin{abstract}
%-------------------------------
Einstein's general relativity predicts that orbital motion of accreted gas approaching a neutron star (NS) in a NS low-mass X-ray binary (LMXB) system occurs on a time scale proportional to the NS mass. Radiation of the gas accounts for most of the observed LMXBs variability. In more than a dozen of sources twin-peak quasi-periodic oscillations (QPOs) have been observed. Inspired by the expected proportionality between periods of orbital motion and NS mass we present a straightforward comparison among these sources. We investigate relations between QPO periods and their ratios and identify characteristic time scales of QPOs associated to individual sources.  {These timescales {are likely determined by} the relative mass of each NS.} {We show that the characteristic time scale of the millisecond pulsar XTE J1807.4-294 is longer than for most other NS LMXBs. Consequently, models of QPOs that consider geodesic orbital frequencies imply that the X-ray pulsars' mass has to be about $50\%$ higher than {the average mass of} other sources. Consideration of other X-ray pulsars indicates that the exceptionality of XTE J1807.4-294 cannot be related to NS magnetic field in any simple manner. We suggest that QPOs observed in this source can help to discriminate between the proposed versions of the NS equation of state.}
% We suggest that QPOs observed in this source challenges NS equation of state.
%-------------------------------
\end{abstract}
%-------------------------------


%-------------------------------
\section{Introduction}\label{intro}
%-------------------------------
Accreting relativistic compact objects provide a unique opportunity to observe effects associated with strong gravity in both black hole (BH) and neutron star (NS) systems \citep[][]{kli:2006,mcc-rem:2003,abr-fra:living}. The latter systems may serve as a good tool for exploration of  supra-dense matter \citep[][]{web,lat:2004}. {A large number of observations of X-ray radiation from NS Low-mass X-ray binaries (LMXBs) has been gathered over the past two decades. These systems exhibit a very complex phenomenology including distinct types of spectral behaviour and temporal variability. From the spectral evolution point of view, NS LMXBs are often classified as atoll- or Z- sources based on the shape of the track which they follow in the so-called colour-colour diagram \citep[e.g.][]{kli:2006}. While Z sources are generally more stable and brighter, atoll sources are weaker and show significant changes in the X-ray luminosity. In both types of sources, the power-density spectrum (PDS) that represents their variability commonly include broad-band noise continuum with descending shape often accompanied by more or less sharp peaks.}

{Some of the sources, the accreting milisecond pulsars (AMXPS), pose very sharp peaks - X-ray pulsations - that are associated to NS rotation and the influence of NS magnetic field on the accreted matter. At present there is neverthless no detailed consensus on the NS magnetic field strength in the accreting LMXBs. There is a general agreement that the surface dipole field at the NS stellar equator reaches values of $B\sim10^7\!-\!10^{10}$G {\citep{muk-etal:2015:}}. More detailed estimates are inferred from miscellaneous hypotheses of NS evolution and it is not even clear whether or not the X-ray pulsations can be linked to magnetic field more strongly than in other sources. In most cases the AMXPS belong among atoll sources that are believed to have magnetic fields that are weaker than for the Z-sources \citep[][]{pat-wat:2012:}.}

Less sharp peaks that often appear in the NS PDS are called quasi-periodic oscillations (QPOs). These are usually distinguished into groups of low- and high- frequency QPOs. The low frequency QPOs in NS sources have frequencies in the range of $1-100$Hz. In the case of Z-sources they have been further classified into horizontal, flaring, and normal branch oscillations (HBO, FBO, and NBO, respectively) depending on the position of the source in the colour-colour diagram. Oscillations with properties similar to HBOs have also been observed in several atoll sources \citep[see][ for a review]{kli:2006}.  The high-frequency (HF) QPOs in the NS sources span a frequency range of tens to more than thousands of hertz. They frequently appear in pairs that are observed simultaneously at frequencies $\nuU>\nuL$. Hence the name twin-peak QPOs by which they are commonly known. Sources that exhibit both X-ray pulsations and twin-peak QPOs are very rare, but they do exist {.\footnote{{So far only three sources have displayed both, a number of twin-peak QPO detections, and strong X-ray pulsations \citep[][]{men-bel:2007:,pat-wat:2012:}. These sources are further considered within our paper - see Table~\ref{table:1} for the list of them.}}}

{The strength of peaks that form the twin-peak QPOs expressed in terms of the fractional root-mean-squared (rms) amplitude, $r$, varies sometimes being as high as $r\sim30\%$. The coherence of the signal may vary as well. There have been reported peaks with quality factor (centroid frequency of the peak divided by its full width at half maximum) of up to $Q=300$ \citep[][]{bar-klu:2005,bar-etal:2005,bar-vau:2012}. Variable frequencies along with high coherence and high amplitude make NS HF QPOs different from the weak HF QPOs that are observed in BH sources \citep[e.g.][]{Rem-etal:2002:, mcc-rem:2006}. Those are associated with having stable frequencies that often form a 3:2 frequency ratio \citep[first noticed by][]{abr-klu:2001},}
%----------------------------
\begin{equation}
%----------------------------
{R\equiv\nu_\U/\nu_\L = 1.5.}
%----------------------------
\end{equation}
%----------------------------
{However, see the works of \citet{bel-etal:2012}, \citet{bel-alt:2013} and \citet{var-rod:2018} where the robustness of the 3:2 ratio is challenged.}

{Robust correlations are observed between the frequencies of twin-peak QPOs.} Each source reveals its specific frequency correlation, $\nu_\U = \nu_\U(\nu_\L)$, although the sources roughly follow a common pattern \citep[][]{psa-etal:1999a,abr-etal:2005:AN,abr-etal:2005:RAG,zha-etal:2006}. {Despite the fact that QPOs have now been observed for more than three decades, the origin of both LF and HF QPOs is still poorly understood.} There is presently no commonly accepted model for either BH or NS HF QPOs. On the other hand, based on several strong arguments, it is generally expected that these oscillations originate in orbital motion in the vicinity of the compact object.

%------------------
\section{Orbital models of QPOs}
%------------------
\label{section:models}

{Various competing models of QPOs have been proposed. It is usually assumed that the QPO excitation occurs within the most luminous accretion region located less than two tens of Schwarzschild radii above the inner edge of the accretion disc. Several models suggest that QPOs are produced by a local motion of accreted inhomogeneities such as blobs or vortices. This subset of {QPO} models includes the so-called relativistic precession (RP) or tidal disruption (TD) model \citep[][]{abr-etal:1992,ste-vie:1998b,ste-vie:1999,cad-etal:2008,kos-etal:2009,bak-etal:2014,kar-etal:2017,ger-etal:2017}.}

Another possibility is to relate the QPOs to a collective motion of the accreted matter, in particular to oscillatory modes of an accretion  disc \cite[][]{wag-etal:2001,rez-etal:2003,abr-etal:2006,Ingram+Done:2010,fragile-blaes:2016}. Such models often work with the idea of oscillations in a slender accretion torus and some sort of resonance between the torus oscillatory modes. An example is the epicyclic resonance (Ep) model proposed by \cite{klu-abr:2001}. Several other models have been discussed by a large group of authors \citep[see, e.g.][]{alp-sha:1985,lam-etal:1985,mil-etal:1998a,psa-etal:1999b,wag:1999,psa-col:2000,wag-etal:2001,abr-klu:2001,klu-abr:2001,kat:2001,tit-ken:2002,abr-etal:2003b,abr-etal:2003c,rez-etal:2003,klu-etal:2004,pet:2005a,zha:2004,zha:2005,bur:2005,tor-etal:2007,kat:2007,kat:2008,stu-etal:2008,cad-etal:2008, kos-etal:2009, ger-etal:2009,muk:2009,stu-etal:2013,stu-etal:2014,torok-etal:2015:MNRAS:,wan-etal:2015:MNRAS:,stu-etal:2015,shi-etal:2009,shi-etal:2018}.

%--------------------------------------
{\subsection{Frequency relation associated to CT model}}
%--------------------------------------

{\cite{tor-etal:2018} have recently found and explored a surprisingly simple analytic relation that reproduces individual correlations for a group of several sources through a single parameter,}
%-------------------------------
\begin{equation}
%-------------------------------
\label{equation:the-one}  
\nuL= \nuU\left(1 - {\mathcal{B}}\sqrt{1 - \left(\nuU/\nu_0\right)^{2/3}}\right),
%-------------------------------
\end{equation}
%-------------------------------
{where $\nu_0$ equals to Keplerian frequency at the innermost stable circular orbit (ISCO), $\nu_0=\nuISCO$, and $\mathcal{B}=0.8$. In the Schwarzschild spacetime, $\nuISCO$ is given solely by the NS mass $M$.} The authors argue that relation (\ref{equation:the-one}) supports the hypothesis of the orbital origin of twin-peak QPOs. They discuss its interpretation in terms of global motion of the accreted fluid.

{For $\mathcal{B}=1$ equation (\ref{equation:the-one}) describes predictions of the RP model, while the same relation provides predictions of a model that assumes an oscillating pressure-supported torus located at the innermost accretion region when $\mathcal{B}=0.8$.\footnote{{\cite{tor-etal:2018} have also investigated data fitting with $\mathcal{B}$ being a free parameter further improving the fits in some sources. They discussed deviations from $\mathcal{B}=0.8$ in terms of non-geodesic effects other than the influence of the torus pressure.}} The torus is assumed to form a cusp by filling up its critical equipotential volume \citep[see][for a further context]{rez-etal:2003,zan-etal:2005,tor-etal:2016:MNRAS,ave-etal:2017,tor-etal:2018}.  We refer to this model as to CT model. \cite{tor-etal:2018} have found that non-rotating NS mass inferred from equation (\ref{equation:the-one}) and $\mathcal{B}=0.8$ does not much vary across the individual sources; there is $M<2M_{\odot}$ except for the outstanding case of the XTE J1807.4-294 milisecond pulsar which is associated to high values of $M$, $M>2.4M_{\odot}$. {In next} we explore this finding and discuss its interpretation and importance in general context of the orbital QPO models.}

%-------------------------------
\section{Relativistic scaling of orbital frequencies and BH QPOs}
%-------------------------------

Characteristic periods $T$ of geodesic orbital motion in strong gravity scale with the compact object mass $M$. Several models of QPOs, to which we in next refer as {standard geodesic orbital (SGO)} models, assume that the observed frequencies, $\nuL$ and $\nuU$, are equal to frequencies of (geodesic) orbital motion represented by the Keplerian frequency of a circular orbit, by the radial and vertical epicyclic frequencies, or by their linear combinations (including the periastron and Lense-Thiring precession frequencies). These frequencies can be written in a general form
%-------------------------------
\begin{equation}
%-------------------------------
\label{equation:uni}
\nu_{\mathrm{i}} = \frac{1}{T_{\mathrm{i}}}= \frac{c^3}{2\pi  G}\frac{1}{M}f_{\mathrm{i}}(x,j, q),
%-------------------------------
\end{equation}
%-------------------------------
where {the orbital radius is rescaled as} $x=r/r_{G}$, $r_{G}=GM/c^2$, $j$ denotes the compact object (BH or NS) rotational parameter (dimensionless angular momentum), $j=cJ/GM^2$, and $q$ stands for quadrupole moment, $q=cQ/GM^3$. {The $f_{\mathrm{i}}$ function which determines given orbital frequency depends on $j$ and $q$, but does not depend on $M$.} For simplicity we from now on assume very compact objects (BH or dense NS), $q\approx j^2$.

In the specific case of the Keplerian frequency we can write the $f$ function as \citep[e.g.][]{bar-etal:1972,abr-etal:2003:HT}
%-------------------------------
\begin{equation}
%-------------------------------
\label{equation:Kepler}
f_{\K} = \frac{1}{x^{3/2}+j}.
%-------------------------------
\end{equation}
%-------------------------------
An often quoted example of $1/M$ scaling of the orbital frequencies is related to the innermost stable circular orbit. The Keplerian orbital frequency at this orbit around a Schwarzschild black hole, $j=0,~x_{\ISCO}=6$, scales as \citep[e.g.][]{wag-klu:1985:}
%-------------------------------
\begin{equation}
%-------------------------------
\label{equation:ISCO}
\nuISCO=\frac{2.2\mathrm{kHz}}{M^{*}},
%-------------------------------
\end{equation}
%-------------------------------
where $M^{*}=M/M_{\sun}$. For an extremely rotating Kerr black hole, $j=1,~x_{\ISCO}=1$, the ISCO frequency is higher and one may write
%-------------------------------
\begin{equation}
%-------------------------------
\nuISCO=\frac{16.2\mathrm{kHz}}{M^{*}}.
%-------------------------------
\end{equation}
%-------------------------------

%-------------------------------
\subsection{Universal scaling of QPO frequencies in BH sources}
%-------------------------------

%--------------------------------------------------------
%----------- FIGURE 1         ---------------------------
%--------------------------------------------------------
%---------------------------------------------------------
\begin{figure}
%---------------------------------------------------------
\begin{center}
%a) \hfill ~~~~~~~~~b)~~~~~~~~~~~~~~~~~~~~~~~~~~~~~~~~~~~~~~c) \hfill ~$\phantom{d}$ 
\includegraphics[width=1\linewidth]{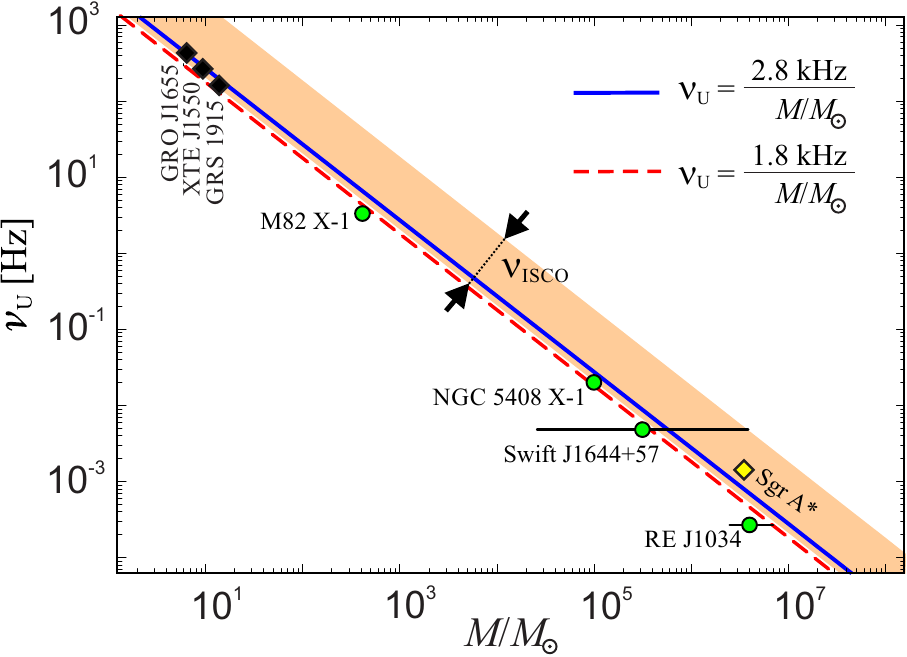}
\end{center}
\caption{Large scaling of BH 3:2 QPO frequencies. The dashed line corresponding to $\nuU=1.8$kHz indicates the RP model prediction for the 3:2 frequency ratio and $j=0$. The coloured region denotes the ISCO frequencies in the range of $j\in[0,~1]$.}
%\caption{a) Large scaling of BH HF QPO frequencies. Dashed line corresponding to $\nuU=1.8$kHz denotes prediction of RP model for the 3:2 frequency ratio and $j=0$. b) Comparison between periods of BH and NS HF QPOs. Dashed vertical line denotes $R=1.5$. Continuous lines indicate predictions of RP model (in the bottom part of the plot, the curves  are spaced every 1.4 in $M^*$). c) Enlarged view of panel b.}
\label{figure:1}
%---------------------------------------------------------
\end{figure}
%---------------------------------------------------------

It has been shown that the 3:2 frequencies observed in Galactic microquasars scale qualitatively in the same way as the above mentioned relations. The best fit of their data by the $1/M$ relation has been determined as \citep[][]{mcc-rem:2003}
%-------------------------------
\begin{equation}
%-------------------------------
\label{equation:bestfit}
%\nuU=\frac{1}{T_\U} = \frac{2.8\mathrm{kHz}}{M^{*}}.
\nuU = \frac{2.8\mathrm{kHz}}{M^{*}}.
%-------------------------------
\end{equation}
%-------------------------------
All Galactic microquasars therefore should have its rotational parameter similar to one another, except when the QPO frequencies are not much affected by its value \citep[e.g.][]{tor-etal:2007:}. For instance, the RP model predicts relation (\ref{equation:bestfit}) for the observed 3:2 frequency ratio for $j\sim0.5$ \citep[e.g.][]{tor-etal:2011}. On the other hand the Ep model predicts $j\sim0.9$ \citep[][]{tor-etal:2005}.

It has been suggested that on a large range of $M$ both the rotational parameter and specific details of a given orbital model are of secondary importance, and the observed frequencies can be used for the estimation of $M$ \citep[][]{abr-etal:2004,tor:2005}. At present there is a growing evidence for such a large range BH QPO frequencies scaling \citep[][]{zho-etal:2015:ApJL}. This is illustrated in Figure \ref{figure:1} which indicates the QPO data and expected BH masses investigated by \cite{Rem-etal:2002:,rem:2004,pas-etal:2014,stro-etal:2007,reis-etal:2012,gie-etal:2008,ach-etal:2004,rem-mc:2006,rei-etal:2014,hua-etal:2013,mil-gul:2011,zhou-etal:2010,gil-etal:2009}. In the Figure \ref{figure:1} we include the $\nuU(M)$ relations associated to Keplerian frequency at ISCO as well as those predicted by the RP model. {Further references and examples of other BH sources are discussed by \citet{gol-etal:2019:} {and \cite{gup-etal:2019}.}}

%--------------------------------------------------------
%----------- FIGURE 2         ---------------------------
%--------------------------------------------------------
%---------------------------------------------------------
\begin{figure*}
a) \hfill ~~~~ b)\hfill $\phantom{c}$
\vspace{-1ex}

\begin{center}
\includegraphics[width=1.\linewidth]{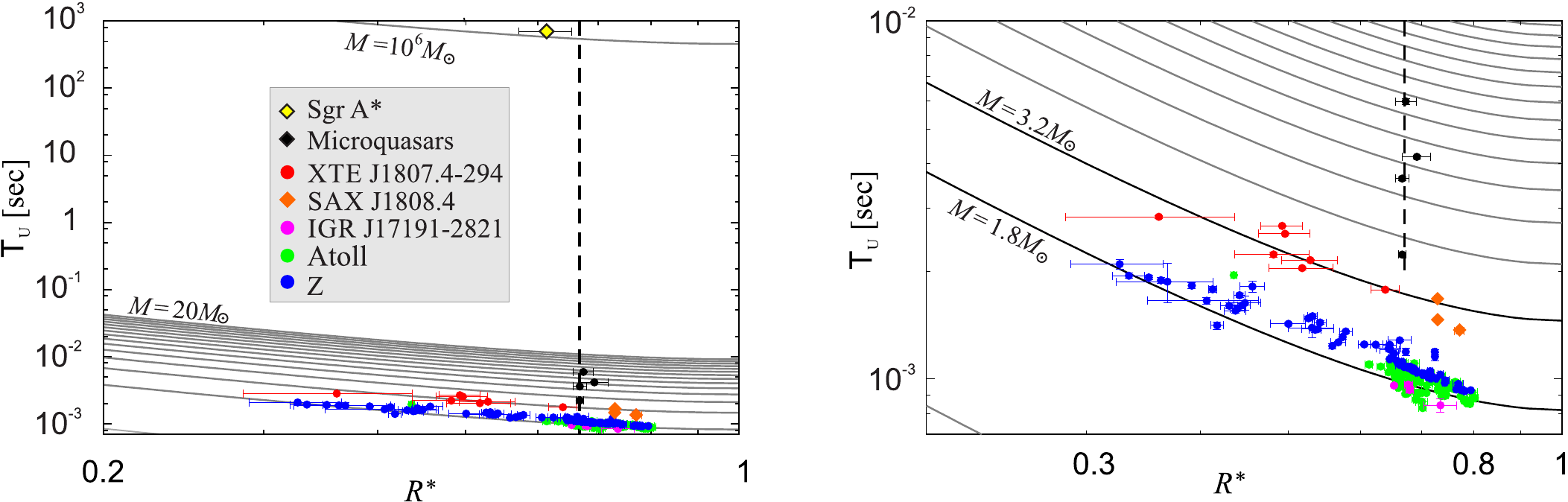}
\end{center}

\caption{a) A comparison between the periods of BH and NS HF QPOs. The dashed vertical line corresponds to $R=1.5$. The continuous lines indicate predictions of the RP model. In the bottom part of the plot the curves  are spaced every 1.4 in $M^*$. b) An enlarged view of panel a.}
\label{figure:2}
\end{figure*}
%---------------------------------------------------------

%-------------------------------
%\section{Comparison between BH, atoll, Z and X-ray pulsar sources}
%\section{Comparing timescales of QPOs in NS sources}
\section{Comparison between % Není between jen mezi dvěmi (AMONG)? !!!!!! ????????????????????????
	     NS sources}
%-------------------------------

{As we mentioned in Section~\ref{section:models}, there are several competing QPO models that are still premature. It still is not clear whether the same model could be applied to both (BH and NS) classes of sources. The fixed frequencies of the weak BH QPOs seem to be in contrast with the variable frequencies of the often strong NS QPOs. It has been suggested within the framework of the SGO models that the fixed frequency ratio, $R=1.5$, observed in the BH sources relates to a specific resonant orbit $x_{3:2}$ \citep[e.g.][]{tor:2005}. Local (e.g the RP model) or global (e.g. the Ep model) oscillations of the accreted matter associated to this orbit have been shown to possibly evoke the observed variations of the flux in the BH sources \citep[][]{bur-etal:2004,sch-ber:2004:,bak-etal:2015}.}

{It has been proposed that physical mechanisms that occur at the NS boundary layer can enhance the observed variations of the flux in such a way that their amplitude is up to one order of magnitude higher than for the BH sources \citep[][]{hor:2005,abr-etal:2007,tor-etal:2016:MNRAS,pas-etal:2017}. The turbulent environment of the NS boundary layer can evoke variations of the QPO excitation radius. {Within a possible unified BH-NS QPO model the presence of boundary layer and absence of event horizont can be responsible for the differences between BH and NS sources.}  One may naturally expect that in such model the $1/M$ scaling of QPO frequencies found in BH sources should be manifested in the NS sources as well. We note that some $1/M$ scaling may be expected in the NS sources even if the QPO mechanisms in the NS and BH sources were different provided that the mechanisms are described by some of the SGO models (that incorporate relativistic scaling of the orbital frequencies).}

%-------------------------------
\subsection{Timescales}
%-------------------------------

{Within SGO class of models,} the observed NS correlations $\nuU(\nuL)$ can be related to a variable orbital radius $x$ \citep[e.g.][]{ste-vie:1998a,ste-vie:1998b}. Assuming equation (\ref{equation:uni}), for a fixed $j$, the ratio of QPO periods, ${R}^{*}$, depends only on $x$ and not on $M$,
%-------------------------------
\begin{equation}
%-------------------------------
\label{equation:ratio}
R^{*}\equiv\frac{T_{\U}}{T_{\L}}=\frac{\nuL}{\nuU}=\frac{1}{R(x)}.
%-------------------------------
\end{equation}
%-------------------------------
Moreover, for several models, $R^{*}$ is a monotonic function of $x$. 

Relations (\ref{equation:uni}) and (\ref{equation:ratio}) imply that, for a fixed $R^*$, higher QPO  periods correspond to a higher compact object mass. Having this motivation in mind, in Figure~\ref{figure:2} we include QPO periods observed in the NS sources. Within the same Figure we also illustrate predictions of the RP model which implies $r\rightarrow \infty$ when $R^*\rightarrow 0$ and $r\rightarrow r_{\ISCO}$ when $R^*\rightarrow 1$. {The NS sources included in Figure~\ref{figure:2} are listed in Table   ~\ref{table:1}}.\footnote{{We note that we do not consider Circinus X-1 in this paper because of the large extension of its $R$ error bars \citep[see Section 5.1 in ][]{tor-etal:2012}.}}

Inspecting Figure~\ref{figure:2} we can find that, although there are some differences \citep[see also][]{wan-etal:2014,wan-etal:2018}, both atoll and Z sources seem to roughly follow a common correlation $T_{\U}(R^{*})$. On the other hand, we can see that the XTE J1807.4-294  milisecond pulsar follows a correlation $T_{\U}(R^{*})$ that is quantitatively different from that of the other displayed atoll and Z-sources {except for {SAX J1808.4-3658}}.

%--------------------------------------------------------
%----------- FIGURE 3         ---------------------------
%--------------------------------------------------------
%---------------------------------------------------------
\begin{figure}
%---------------------------------------------------------
\begin{center}
%a) \hfill ~~~~~~~~~b)~~~~~~~~~~~~~~~~~~~~~~~~~~~~~~~~~~~~~~c) \hfill ~$\phantom{d}$ 
\includegraphics[width=.93\linewidth]{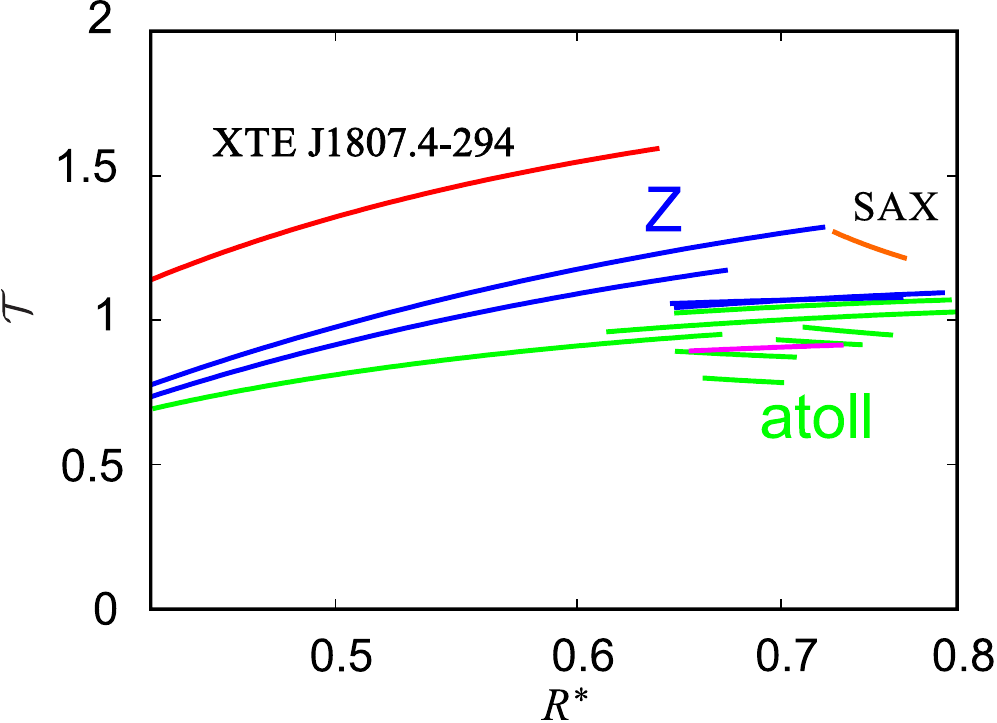}
\end{center}
\caption{{The $\mathcal{T}(R^{*})$ functions obtained for the individual sources and $\bar{T}_{\U}(R^{*})$ inferred from the common fit of all datapoints. {The colour coding is the same as in Figure~\ref{figure:2}.}}}
\label{figure:3}
%---------------------------------------------------------
\end{figure}
%---------------------------------------------------------

%--------------------------------------------------------
%----------- FIGURE 4        ---------------------------
%--------------------------------------------------------
%---------------------------------------------------------
\begin{figure*}
\begin{center}
a) \hfill ~~~~~~b) \hfill ~$\phantom{d}$\\
\includegraphics[width=.9\linewidth]{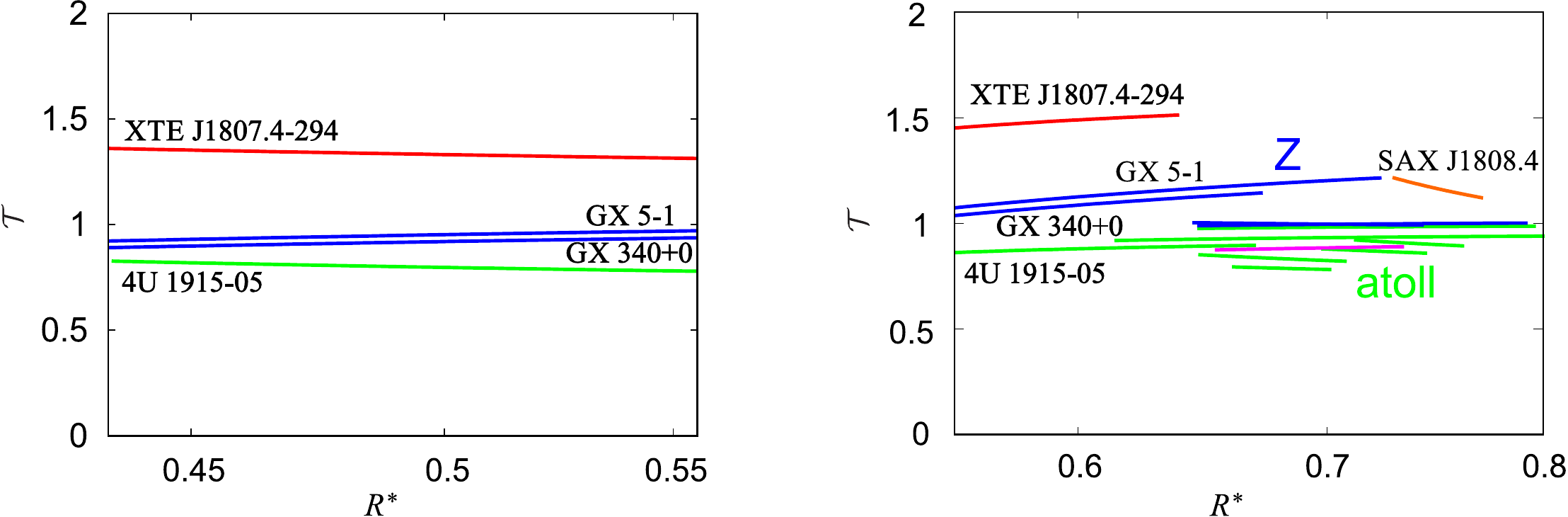}
\end{center}
\caption{{The $\mathcal{T}(R^{*})$ functions obtained for the individual sources and $\bar{T}_{\U}(R^{*})$ given by relation (\ref{equation:bar}).} {{The colour coding is the same as in Figure~\ref{figure:2}.} a) Results for the $R^{*}_{\mathrm{I}}$ interval. b) Results for the $R^{*}_{\mathrm{II}}$ interval.}}
\label{figure:4}
\end{figure*}
%---------------------------------------------------------

%---------------------
\subsection{{Relative periods}}
%---------------------

We attempt to quantify the difference between XTE J1807.4-294 and other sources. We interpolate individual data of each of the {14} sources using relations (\ref{equation:the-one}) and (\ref{equation:ISCO}) with best fitting coefficients $\mathcal{M}$ and $\mathcal{B}$. In this way we obtain continuous correlations $T_{\U}(R^{*})$ that well match the observational data.

{We performed a common fit of data of all sources using relation (\ref{equation:the-one}). The fit {($\mathcal{M}=1.64\pm{0.02}$, $\mathcal{B}=0.67\pm{0.01}$)} defines an averaged correlation $\bar{T}_{\U}(R^{*})$ that can be used to explore the behaviour of relative periods}
%-------------------------------
\begin{equation}
%-------------------------------
\label{equation:tau}
\mathcal{T}\equiv{T}_{\U}/\bar{T}_{\U},
%-------------------------------
\end{equation}
%-------------------------------
{where individual continuous correlations $T_{\U}(R^{*})$ interpolate individual data of each of the {14} sources using relations (\ref{equation:the-one}) and (\ref{equation:ISCO}) with best fitting coefficients $M$ and $\mathcal{B}$. The relative periods $\mathcal{T}(R^*)$ obtained for each source are displayed in Figure~\ref{figure:3}. The Figure shows that the longest relative QPO periods of values close to $\mathcal{T}\sim1.5$ are reached for the XTE J1807.4-294 pulsar.}

%-------------------------------
%\subsection{Mass}
\section{{Scaling of relative periods with NS mass}}
%-------------------------------

For a given source, neglecting the effects of NS rotation and assuming only SGO models, $\mathcal{T}$ has to be a constant which only depends on $M$,
%-------------------------------
\begin{equation}
%-------------------------------
\label{equation:tau2}
\mathcal{T}=\mathcal{T}_{0}(M).
%-------------------------------
\end{equation}
%-------------------------------
For the whole class of SGO models, there is 
%-------------------------------
\begin{equation}
%-------------------------------
\label{equation:tau3}
M =\tau\,\mathcal{T}_{0}\mathcal{F},
%-------------------------------
\end{equation}
%-------------------------------
where $\tau$ is the absolute period corresponding to $\mathcal{T}=1$ and $\mathcal{F}$ is a factor specific for a given model. Consequently, a higher value of $\mathcal{T}$ corresponds to a higher $M$.

%----------------------------
{\subsection{Individual sources' behaviour}}
%----------------------------

{The curves drawn in Figure~\ref{figure:3} clearly differ from constant functions. This can be an artefact of the application of common fit of data of all sources that was used to explore the behaviour of relative periods. In order to avoid bias connected to the non-uniform coverage of the QPO data along the large range of $R^*$ we divide the examined data into two intervals, $R^{*}_{\mathrm{I}} \in [0.43,~0.56]$ and $R^{*}_{\mathrm{II}} \in [0.56,~0.8]$. For the normalization of periods within the group of $\mathrm{n}$ sources we then consider an averaged correlation $\bar{T}_{\U}(R  ^{*})$,}
%-------------------------------
\begin{equation}
%-------------------------------
\label{equation:bar}
\bar{T}_{\U}(R^{*})=\sum_{\mathrm{No.}=1}^{\mathrm{n}}T_{\U}^{\mathrm{No.}}(R^{*})/{\mathrm{n}},
%-------------------------------
\end{equation}
%-------------------------------
{instead of those given by the common fit. We note that there are only three datapoints available for {SAX J1808.4-3658}. {These datapoints cover only a small part of the relevant interval and we exclude them from the calculation of $\bar{T}$.} The $\mathcal{T}(R^{*})$ functions obtained for the two intervals of $R^*$ are shown in Figure~\ref{figure:4}.

%----------------------------
{\subsection{Distribution of $\mathcal{T}_{0}$}}
%----------------------------

{Figure~\ref{figure:4} provides a rough quantitative comparison between different sources. Through approximation of the curves displayed within the $R^*_{II}$ interval by straight lines, one can obtain a distribution of $\mathcal{T}_{0}$ values. This specific distribution is illustrated in Figure~\ref{figure:5}a. The uncertainties related to individual values of $\mathcal{T}_{0}$ within this distribution depend on very specific assumptions and could easily be underestimated.}

{In order to determine the value of $\mathcal{T}_0$ and its uncertainty for each source in a rigorous and more robust way we analyze a full sample of data, as well as the individual sources separately, both across the whole range of $R^*$. Based on the fits obtained, we estimate mean values of $\mathcal{T}_{0}$ that determine the relative mass of each source. We extract information on their uncertainties by performing Monte-Carlo simulations that provide 2-dimensional distributions of the fitting parameters $\mathcal{M}$ and $\mathcal{B}$. The obtained results are given in Table~\ref{table:1} and illustrated in Figure~\ref{figure:5}b.}

% In NS sources it is usually expected that $j\in[0,~0.3]$ and $\nuISCO$ does not much differ from $\nuISCO(j=0)$.

%--------------------------------------------------------
%----------- FIGURE 5         ---------------------------
%--------------------------------------------------------
%---------------------------------------------------------
\begin{figure*}
%---------------------------------------------------------
\begin{center}

\includegraphics[width=.75\linewidth]{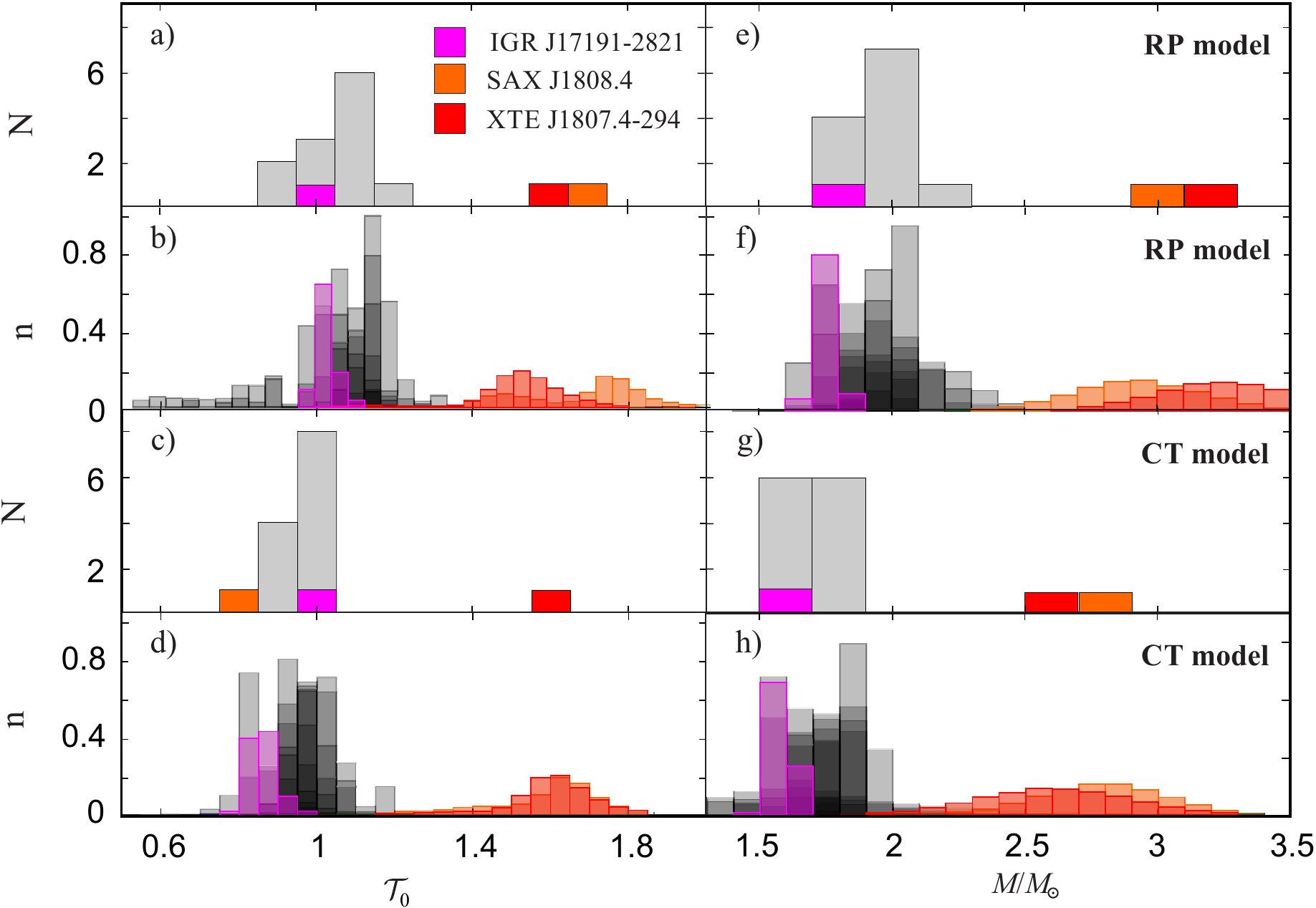}
\end{center}

\caption {{a) Distribution of $\mathcal{T}_{0}$ values associated to curves displayed in Figure~\ref{figure:4}b. The quantity N denotes the number of occurences. b) Distribution of the estimated $\mathcal{T}_{0}$ values associated to individual sources drawn from Monte-Carlo simulations assuming 2-dimensional distributions in best fitting parameters given by relation (\ref{equation:the-one}). The quantity n denotes the relative number of occurences. c) The same as in panel a), but made for the linear formula instead of relation (\ref{equation:the-one}). d) The same as in panel b), but made for the linear formula instead of relation (\ref{equation:the-one}). e) Distribution of NS mass implied by fitting the QPO data of the individual sources by the RP model for $j=0$. f) A detailed estimation of uncertainties in NS mass drawn for the RP model, $j=0$, and each source. g) Distribution of NS mass implied by fitting the QPO data of the individual sources by the CT model for $j=0$. h) A detailed estimation of uncertainties in NS mass drawn for the CT model, $j=0$, and each source.}}
\label{figure:5}
%---------------------------------------------------------
\end{figure*}
%---------------------------------------------------------

\bigskip
%-------------------------------
\section{Discussion and conclusions}
%-------------------------------

{The distribution of $\mathcal{T}_{0}$ drawn in Figures~\ref{figure:5}a,b nicely illustrates the exceptionality of XTE J1807.4-294 - {its characteristic QPO timescale is of a factor of $50\%$ longer than the average timescale of other sources.}} For any SGO model and $j=0$ our finding implies that, compared to others, this source has a very high mass. {We note that our conclusion is rather robust and does not depend on the exact form of formula (\ref{equation:the-one}) that we use to obtain the best data interpolation. For instance, when the formula is replaced by a simple linear term, $\nuU=\mathrm{a}\nuL+\mathrm{b}$, a very similar result is achieved, see {Figures~\ref{figure:5}c,d.}}
 
{An overall inspection of Figures~\ref{figure:5}a-d clearly supports our conclusion. We further illustrate this conclusion} and display the NS mass distribution inferred from the RP and CT models and the XTE J1807.4-294 QPO data in {Figures~\ref{figure:5}e-h} that follows the results of \cite{tor-etal:2016:MNRAS}. It shows that for both QPO models and $j=0$ the inferred mass of the X-ray pulsar is indeed very high compared to other sources.}

%-------------------------------
%\subsection{Rotating NSs}
\subsection{{NS rotation and equation of state}}
%-------------------------------
\label{section:EoS}

\begin{figure*}
%---------------------------------------------------------
\begin{center}
a) \hfill ~~~~~~~~~b) \hfill ~$\phantom{d}$ 
\includegraphics[width=1\linewidth]{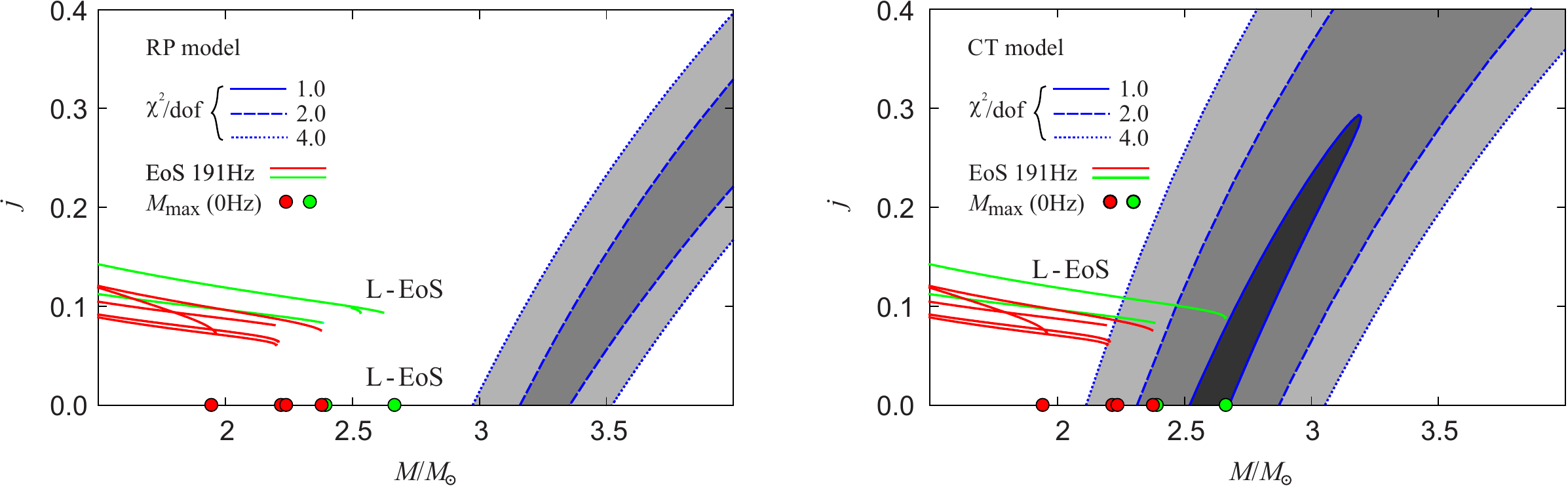}
\end{center}
\caption{The mass--angular-momentum contours obtained from fitting of datapoints of XTE J1807.4-294 by QPO models vs. mass-–angular-momentum relations predicted by models of rotating NSs. These are drawn for several NS EoS and spin 191Hz inferred from the X-ray burst measurements. {The red colour corresponds to EoS assumed by \citet{tor-etal:2016:MNRAS}. Restrictions given by models assuming non-rotating NSs are also indicated. See Section~\ref{section:EoS} for details.} a) RP model. b) CT model.}

\label{figure:6}
%---------------------------------------------------------
\end{figure*}

In Figure~\ref{figure:6} we illustrate NS mass and angular momentum as implied by fitting of the  {XTE J1807.4-294} QPO data by the RP and CT models for $j\geq 0$. Within the same Figure we include NS mass constraints following from several NS equations of state (EoS). {These are namely EoS considered by \cite{tor-etal:2016:MNRAS} - SLy 4, APR, AU-WFF1, UU-WFF2 and WS-WFF3 \citep{wir-etal:1988,ste-fri:1995,akm-etal:1998,rik-etal:2003}, and two more EoS - L, l \citep{arn+bow:1977,urb-etal:2010:aca}.} In this Figure we assume the NS rotational frequency of $191$Hz reported by \cite{lin-etal:2005,bou-lam:2008}. The calculations were performed following the approach of \cite{har-tho:1968}, \cite{cha-mil:1974}, \cite{mil:1977}, \cite{urb-etal:2013}, \cite{tor-etal:2012} and \cite{tor-etal:2016:MNRAS}.

Inspecting Figure~\ref{figure:6} we can see that the non-rotating NS mass implied by the RP model is {rather} high compared to maximal non-rotating NS mass allowed by the assumed EoS. This is the case also for the CT model, even though this model implies considerably lower $M$. On the top of that, the estimated NS mass increases when NS rotation is taken into account. We note that such behavior is common for most SGO models \citep{tor-etal:2016:ApJ}. For this reason it may be very difficult to match the long QPO periods in {XTE J1807.4-294} when realistic models of rotating NS and SGO models are considered simultaneously.

{Our findings on the SGO models and NS EoS are well illustrated by the example of the CT model shown in Figure~\ref{figure:6}b. EoS that allows for a very high NS mass is clearly required. In this sense we can state that the QPOs observed in {XTE J1807.4-294} challenge the NS EoS.}

%---------------------------------------------------------
%  TABLE 1
%---------------------------------------------------------
\begin{table}
%---------------------------------------------------------

\caption{{List of sources (A - atoll, Z - Z, P - AMXP) and obtained values of $\mathcal{T}_{0}$. {The uncertainties of $\mathcal{T}_{0}$ correspond to standard errors given by $\mathcal{T}_{0}$ distributions shown in Figure~\ref{figure:5}b.}  References: {(1)--(3), (9) -- (11)} - \citet{bar-etal:2005, bar-etal:2005:b,bar-etal:2006}, {(4)} - \citet{boi-etal:2000:}, {(5)} - \citet{alt-etal:2010:},  {(6)} -  \citet{Hom-etal:2002:}, {(7)} - \citet{bou-etal:2006:},  {(8)} - \citet{lin-etal:2005}, {(12)} - \citet{Jon-etal:2000:}, {(13)} - \citet{jon-etal:2002:},  {(14)} - \citet{bul-kli:2015}.  }}
\vspace{-3ex}
 \label{table:1}

\begin{center}
\renewcommand{\arraystretch}{1.0}
{\begin{tabular}{cll}\hline \hline
Source& Name   & $\mathcal{T}_{0}$  \\
No./type\\
1/A & 4U 1608-52  &{${1.11^{\pm 0.01}}$}  \\ 
2/A & 4U 1636-53  & {${1.05^{\pm 0.01}}$ } \\ 
3/A & 4U 1735-44  & {${1.06^{\pm 0.03}}$}  \\ 
4/A & 4U 1915-05  & {${0.99^{\pm 0.03}}$}\\ 
5/A-P & IGR J17191-2821  & {${1.0^{\pm 0.07}}$}  \\ 
6/Z & GX 17+2  & {${1.17^{\pm 0.03}}$}    \\ 
7/Z & Sco X-1  & {${1.12^{\pm 0.01}}$}   \\ 
8/A-P & XTE J1807.4-294  & {${1.58^{\pm 0.20}}$} \\
9/A & 4U  1728-34  & {${0.96^{\pm 0.03}}$}  \\ 
10/A & 4U 0614+09  & {${1.06^{\pm 0.03}}$}  \\ 
11/A & 4U 1820-30  & {${1.11^{\pm 0.03}}$} \\ 
12/Z & GX 340+0  &  {${0.90^{\pm 0.16}}$}   \\ 
13/Z & GX 5-1  & {${0.91^{\pm 0.26}}$} \\ 
14/P &  {SAX J1808.4-3658}  &   {${1.80^{\pm 0.94}}$} \\ 
%16/Z & Circinus X-1  &  ${(0.42^{\pm })}$  \\ 
\hline
%\multicolumn{8}{l}{\parbox[t]{16cm}{$^a$ We note that, in accordance with relation (\ref{equation:the-one}),  the $\mathcal{M}$   parameter  is expressed in  units of Hz$^{-1}$, but its value can be interpreted as the non-rotating NS mass in solar mass units - see equation (\ref{equation:the-other-one}).  Computation of uncertainties in $\mathcal{M}$ is performed in the same way as in \citet{tor-etal:2012}.}} \\
%\multicolumn{8}{l}{\parbox[t]{16cm}{$^b${{The observed frequencies extend below the expected range of physical applicability of the cusp torus model discussed by \citet{tor-etal:2016:MNRAS}.}}}} \\
%\multicolumn{8}{l}{\parbox[t]{16cm}{$^c$ {There is no reliable match by relation (\ref{equation:the-one}). The $\mathcal{M}$ parameter cannot be evaluated.}}}
%\multicolumn{4}{l}{\parbox[t]{17cm}{$^a$ A\,-\,atoll, Z\,-\,Z, P\,-\,pulsar. %Computation of uncertainties in $\mathcal{M}$ is performed in the same way as in \citet{tor-etal:2012}.
 %$^b$ The observed frequencies extend below the expected range of physical applicability of CT model discussed by \citet{tor-etal:2016:MNRAS}.
 %$^d$ {There is no reliable match by relation (\ref{equation:the-one}). The $\mathcal{M}$ parameter cannot be evaluated.}
%}}
\end{tabular}}
\end{center}
%------------------------------------------------------------
\end{table}
%------------------------------------------------------------

%-------------------------------
\subsection{Magnetic field}
%-------------------------------

{As apparent from Figure~\ref{figure:2}b datapoints of  {SAX J1808.4-3658} likely follow the same quantitative trend as datapoints of XTE J1807.4-294. One may speculate that the origin of very high $\mathcal{T}_{0}$ relies in a relatively strong pulsar magnetic field. This suggestion is in agreement with the scenario in which the magnetic field increases the gap between the accretion disc inner edge and the NS surface making the characteristic time scale of the orbital motion longer \citep[see also][]{gho-lam:1978,cam:1992,bak-etal:2010,bak-etal:2012,hab-etal:2018}. Having said that, it is worth mentioning that datapoints of another {AMXPs}, IGR J17191-2821, follow the other sources. The exceptionality of the two sources therefore cannot be explained in terms of their magnetic field {, at least not in a straightforward way}.}

%-------------------------------
\section*{Acknowledgments}
%-------------------------------

We would like to acknowledge the Czech Science Foundation grant No. 17-16287S, the INTER-EXCELLENCE project No. LTI17018 supporting collaboration between the Silesian University in Opava and Astronomical Institute in Prague, and internal grants of the Silesian University in Opava, SGS/14,15/2016. We are grateful to Marek Abramowicz and Ji\v{r}\'{\i} Hor\'{a}k for useful discussions. Furthermore we would like to acknowledge the hospitality of the University of Oxford and the Astronomical Observatory in Rome. {Last but not least, we express our sincere thanks to concierges of Ml\'{y}nsk\'{a} hotel in Uhersk\'{e} Hradi\v{s}t\v{e} for their kind help and participation in organizing frequent workshops of the Silesian university and the Astronomical institute.}
% Dodat inter transfer nebo něco tkového ? !!!!!

%--------------------------------------------------------
%----------- FIGURE 6        ---------------------------
%--------------------------------------------------------
%---------------------------------------------------------
%---------------------------------------------------------

\bibliographystyle{mn2e}
\renewcommand{\baselinestretch}{.87}
\bibliography{reference}

\label{lastpage}

\end{document}